\documentstyle[aps,prl,preprint]{revtex}
\begin{document}
\bibliographystyle{unsrt}

\draft
\title{Acoustic breathers in two-dimensional lattices}
\author{S. Flach$^1$, K. Kladko$^1$ and S. Takeno$^2$}
\address{$^1$ Max-Planck-Institute for Physics of Complex Systems, Bayreuther
Str. 40 H.16, D-01187 Dresden, Germany \\
$^2$ Department of Information Sciences, Osaka Institute of Technology,
1-79-1 Kitayama, Hirakata, Osaka 573-01 Japan}
\date{\today}
\maketitle
\begin{abstract}
The existence of breathers ( 
time-periodic and spatially localized lattice vibrations)
is well established for i) systems without acoustic phonon
branches and ii) systems with acoustic phonons, but
also with additional symmetries preventing the occurence of strains
(dc terms) in the breather solution.
The case of coexistence of strains and acoustic phonon branches
is solved (for simple models) only for one-dimensional lattices.

We calculate breather solutions for a two-dimensional lattice
with one acoustic phonon branch. We start from the easy-to-handle
case of a system with homogeneous (anharmonic) interaction potentials.
We then easily continue the zero-strain breather solution into the model sector
with additional  quadratic and cubic potential terms with the help
of a generalized Newton method. The lattice size is $70\times 70$.
The breather continues to exist, but is dressed with a strain field.
In contrast to the ac breather components, which decay exponentially
in space, the strain field (which has dipole symmetry) should decay
like $1/r^a\;\;,\;\;a=2$. On our rather small lattice we find an exponent 
$a\approx 1.85$. 
\end{abstract}


The understanding of dynamical localization in classical spatially extended
and ordered systems experienced recent considerable progress. Specifically
time-periodic and spatially localized solutions of the classical equations
of motion exist, which are called (discrete) breathers, or intrinsic
localized modes \cite{st88}. The attribute discrete stands for the discreteness of
the system, i.e. instead of field equations one typically considers the
dynamics of degrees of freedom ordered on a spatial lattice. As already
mentioned, the considered systems are spatially ordered, i.e. the lattice
Hamiltonian is invariant under discrete translations in space. The discreteness
of the system produces a cutoff in the wavelength of extended states, and
thus yields a finite upper bound on the spectrum of eigenfrequencies $\Omega_q$
(phonon band) of
small-amplitude plane waves (we assume that for small amplitudes 
the Hamiltonian is in leading order usually a quadratic form of the degrees of 
freedom). If now the equations of motion contain nonlinear terms, the
nonlinearity will in general allow to tune frequencies of periodic orbits
outside of the phonon band, and if all multiples of a given frequency
are outside the phonon band too, there seems to be no further barrier
preventing spatial localization (for a review see \cite{sa96},\cite{fw97}).

The existence of discrete breathers has been so far prooven for
i) weakly coupled anharmonic oscillators \cite{ma94},\cite{db96}
ordered on a lattice of any dimension, and ii) chains of particles 
with nearest neighbour interaction whose potential is a homogeneous
function $\sim z^{2m}$ with $m=2,3,...$ \cite{sf95-pre-1}. 
While the first case starts from the trivial limiting case of noninteracting
oscillators, the second one uses the possibility of space-time separation
(due to the homogeneity property) to reduce the consideration to a 
two-dimensional map. In the first case the phonon band is degenerated
in a nonzero frequency value and can grow upon continuation, keeping
its optical property (i.e. no conservation of total mechanical momentum).
In the second case the phonon band is degenerated in the zero frequency
value, so formally it is an acoustic band (total mechanical momentum
is conserved) but its width is zero.

As already mentioned, the breather frequency $\Omega_b$ should fulfill
a nonresonance condition $n\Omega_b \neq \Omega_q$ for all integer $
n=0,\pm1,\pm2,...$. This is necessary in general 
in order to have spatial localization
of the corresponding Fourier mode \cite{sf94}. In the abovementioned case of weakly
coupled oscillators a proper choice of the breather frequency always
ensures nonresonance. In the case of homogeneous interaction potentials
the symmetry of the potential $\Phi(z)=\Phi(-z)$ is found also in the breather
solution, which implies that only odd Fourier components are present in
the breather solution. Thus the dc component ($0\times \Omega_b=0$) which
is in resonance with the mentioned degenerated phonon band is strictly
zero, and the resonance is harmless.

It is a widespread expectation that breathers play an important role in the
dynamics of anharmonic crystals \cite{sp95}. Since any crystal has acoustic phonon
branches, and the interparticle interaction potentials are not symmetric
around their minimum, one has to face the fact that any breather will
be accompanied by a strain field (gradient of the dc component of
the breather) and that the resonance of the dc component with the 
acoustic phonon branches has to be incorporated into the consideration.

If any nonzero multiple of $\Omega_b$ resonates even with an edge of a
phonon band, this leads either to the vanishing of the whole breather,
or to a delocalization of the breather and to a divergence of its 
energy \cite{fkm96}. The resonance of the dc component
to be considered here is special -
it resonates with a Goldstone mode, and one can expect the resonance
to be not as destructive to the breather as any resonance at nonzero 
frequency. From the theory of elastic defects \cite{LLVII} we know
the characteristic feature of the strain decay to be algebraic
in the distance (from the defect center). The exponent is only depending
on the dimension of the system and on the symmetry character of the
defect (monopole, dipole etc), but independent on the defect strength. 

This independence of the exponent on the defect strength implies, that
if acoustic breathers (breathers with dc components in the presence
of acoustic phonon branches) do exist, there will be no parameter limit
in which their spatial decay becomes infinitely large. Instead the
strain will always decay algebraically , only its amplitude can be
varied. 

At this stage it is appropriate to fix the class of Hamiltonians
to be considered further. We will treat the simplest case of 
hypercubic lattices with one degree of freedom per lattice site and
nearest neighbour interaction, which can be considered as generalized
Fermi-Pasta-Ulam (FPU) systems:
\begin{equation}
H=\sum_l \left[ \frac{1}{2}P^2_l + \sum_{l' \in {\rm DNN} }
\Phi(X_l - X_{l'})  \right] \;\;. \label{1}
\end{equation}
Here $P_l$ and $X_l$ are canonically conjugated scalar momenta and
displacements of a particle at lattice site $l$. Note that depending
on the lattice dimension $d$ the lattice site label $l$ is a $d$-component
vector with integer components. The inner sum in (\ref{1}) goes over
all {\sl directed nearest neighbours}, e.g. for $d=1$ and $l=n$ 
we sum over $l'=n+1$, for $d=2$ and $l=(n,m)$ 
we sum over $l'=\{ (n+1,m);(n,m+1)\}$ etc.
The interaction potential $\Phi(z)$ is given by 
\begin{equation}
\Phi(z)=\frac{1}{2}\phi_2 z^2 + \frac{1}{3}\phi_3 z^3 
+ \frac{1}{4} z^4 \;\;, \label{2}
\end{equation}
which turns out to be generic enough for the purposes discussed below.

Breathers for such a system can be represented in the form
\begin{equation}
X_l(t)=\sum_{k=-\infty}^{+\infty} A_{kl} {\rm e}^{{\rm i}k\Omega_b t}
\;\;. \label{3}
\end{equation}
We will restrict ourselves to solutions invariant under time reversal,
so that all $A_{kl}=A_{-k,l}$ are real. The spatial localization
property of (\ref{3}) implies $A_{k,|l|\rightarrow \infty}
\rightarrow 0$ for $k\neq 0$ and $A_{0,|l|\rightarrow \infty}
\rightarrow {\rm const.}$. 
The dc component of the breather is given by $A_{0l}$.

So far we know only about results for one-dimensional lattices.
Lots of numerical work exist, which show that the acoustic breather
seems to exist as a solution to finite energy 
\cite{bks93},\cite{hsx93},\cite{kbs93}. Its pecularity is that
the dc component of the breather versus lattice site number 
has a kink shape $A_{0,l \rightarrow \pm \infty}
\rightarrow \pm {\rm const.}$ for free boundaries. 
For periodic boundary conditions one would find a linear decay of $A_{0l}$
far from the breather, but the gradient of the dc components (the strain)
is inverse proportional to the size of the chain, so that in the 
limit of an infinite chain the result is again a constant for the dc 
component (zero strain).
An analytical proof has been given recently by
Spicci, Livi and MacKay \cite{lsm97}. The proof considers a diatomic
chain with assymetric interaction potential (note that the corresponding
Hamiltonian differs from (\ref{1}) in that one has to introduce
an additional parameter $1/M\neq 1$ in front of each kinetic energy
term for say all even lattice site indices). The breather is continued
from the limit of zero mass ratio (heavy masses are infinitely heavy).
The problem of resonance with the Goldstone mode is solved by 
coordinate transformation and by imposing a strain field of compact
support. This means that the dc displacements at this limit are given
by a step-like kink. 
The breather is then continued into a sector of the Hamiltonian
with nonzero mass ratio.

The reader might think that we are contradicting ourselves with the 
previous paragraph and the above statements about the algebraic decay
being independent of the breather parameters. 
Let us explain why that is not so. Suppose that
a breather exists, which creates some strain field. The dc displacements
$A_{0l}$ will have some dependence on the lattice site vector $l$.
The strain $E_l$ is given by the lattice gradient of $A_{0l}$.
The far field energy stored is given by the integral over the squared strain.
Assuming that the strain does decay algebraically, we can use continuum
theory far from the breather. The corresponding equation is equivalent
to the electrostatic equations in $d$ dimensions. Consider $d=1$. A monopole
far
field will yield $E=c\neq 0$ and the corresponding energy diverges. Also
in this case the potential $A_{0l} = {\rm sgn}(l)a +cl$. 
This is clearly not what
was observed for acoustic breathers in 1d. A dipole far field instead will
yield $E=0$, $A_{0l} = {\rm sgn}(l)a$, and the  
energy is finite. This is the situation observed. So the known acoustic
breather solutions are accompanied by a dipole strain field. Already
the demand that the acoustic breather is a solution to finite energy 
limits the strain fields to dipole or higher order multipole 
symmetries. In this special case
the potential $A_{0l}$ is constant far away from the breather, so
the corresponding exponent of the algebraic decay is simply zero.
That is the reason why the analytical proof of existence can go through,
because a kink-like field for $A_{0l}$ can have the limiting form
of a step function, which is precisely the case for the limit of zero mass
ratio (see above).

For $d=2$ (square lattice) 
the situation is the following. A monopole will generate
a strain $E\sim 1/l$ and a potential $A_{0l} \sim {\rm ln}(l)$. 
The energy of such a field diverges. If we search for acoustic breathers
with finite energy, we would have to exclude a monopole field.
A dipole generates a strain $E\sim 1/l^2$ (we skip direction dependences 
here) and a potential $A_{0l} \sim 1/l$. The energy for this field
is finite. In any case the predicted exponents of the algebraic decay
are nonzero, and no simple limit exists, which makes the strain to be of
compact support. So already at this stage it is clear, that existence
proofs of acoustic breathers in two-dimensional systems are much more
complicated than for $d=1$.

Notice that for $d=3$ (cubic lattice) 
a monopole generates $E\sim 1/l^2$ and the
energy of this field is finite.

To answer the question 'to be or not to be' we will present numerical
calculations of acoustic breathers of (\ref{1}) for $d=2$. The
results show up to numerical accuracy that acoustic breathers exist
on finite lattices with free boundaries. 
The symmetry and spatial decay properties are in
accord with the expectations given above. The maximum lattice size
is $70\times 70$, but we observed no profound size effects on the
existence and symmetry of the acoustic breather when considering smaller
systems. The only size effect (to be expected) is observed even for the
largest systems with respect to the algebraic decay properties.

We start with $\phi_2=\phi_3=0$. In this case $\Phi(z)=\Phi(-z)$,
so $A_{kl}=0$ for $k=2m$ and $m$ integer. In particular no dc components
are present. Furthermore, due to the degeneracy of the phonon band into
a single number the breathers will be localized in space stronger than
exponentially. Due to the homogeneity of the interaction potential 
we can separate time and space $X_l(t)=U_l G(t)$. The master function
$G(t)$ satisfies the differential equation $\ddot{G}=-G^3$, and the
spatial amplitudes $U_l$ are given by the extrema of a function $S(
\{U_l\})$, i.e. $\partial S/\partial U_l = 0$ :
\begin{equation}
S=\sum_l \left[ \frac{1}{2}U_l^2 -\frac{1}{4}\sum_{l' \in {\rm DNN}}
(U_l-U_{l'})^4 \right ]\;\;. \label{4}
\end{equation}
The  function $S$ has a local minimum at $\{U_l=0\}$. For large values
of the variables $U_l$ it will diverge to $-\infty$ with the fourth
power of the distance from $\{U_l=0\}$ with the exception
of some nongeneric directions in the space of $\{ U_l\}$, in which
$S$ will continue to increase with the second power of the distance from
$\{U_l=0\}$.
Thus all nontrivial extrema of $S$ are saddle points, which are located
on some rim surrounding the point $\{U_l=0\}$.

The search strategy is thus to define a certain initial direction in
$\{ U_l\}$, to find the rim, and then to minimize $S$ staying on the rim.
The procedure is very fast, because localized solutions decay in space
faster than exponentially. The full solution is obtained by multiplying
the found eigenvector for $\{ U_l\}$ with the
time periodic master function $G(t)$, which
can have any period.

After we find a certain solution for $\phi_2=\phi_3=0$ and choose
a certain period $T_b=2\pi / \Omega_b$ for $G(t)$, in the second
step we switch on $\phi_2=\phi_3=0.01$. With the help of a generalized
Newton method (see e.g. \cite{ma96}) we are searching for a
periodic orbit with the same period $T_b$ closely nearby to the starting
solution in phase space. We start with all velocities set to zero,
i.e. with the time point when $\dot{G}(t)=0$. If we find a new
periodic orbit, after time $T_b$ all velocities are zero again, so
in the Newton algorithm we use only the displacement variables $X_l$.
A periodic orbit is said to be found if
\begin{equation}
\sqrt{\sum_l \left[ X_l(t=0)-X_l(t=T_b) \right]^2 } < 10^{-8}
\;\;. \label{5}
\end{equation}
The maximum size of the square lattice $N\times N$ with $N=70$ comes
from the circumstance, that the rank of the Newton matrix is $N^2$,
and the operative memory size needed for calculation with double
precision is $8N^4$ byte. 

The numerical results shown below apply to the abovementioned
initial vector in $\{ U_l\}$ space for which all $U_l$ are zero
except one elementary plaquette of four lattice sites 
on which $|U_l=1|$ and the signs are alternating between nearest neighbours.
We obtained similar results with an initial vector where all $U_l$ are zero
except for one single lattice site where $U_l=1$.

As already mentioned, the Newton search algorithm successfully produced
solutions in all cases considered. 
The ac components of the found solution decay exponentially in space
and essentially vanish at a distance of 5-7 lattice constants from
the center of the breather.
In fig.1 we show the dc displacements
of one solution with a period obtained by initial conditions $G(t=0)=1$
and $\dot{G}(t=0)$ for the master function $G(t)$. We do observe
dipole symmetry of the dc field. In fig.2 a zoom of the center of
the dc field is shown.

Let us turn to the strain. In fig.3 we show the absolute values
of the strain field of the found acoustic breather. 
To analyze the spatial behaviour of the strain, we plot in fig.4
the variation of the absolute values of the strain along the two
diagonals, as in those directions we have the largest distance
and can hope that the boundary effects are supressed in some bulk
region. The results depend on the choice of the diagonal. The diagonal
which is directed along the dipole moment gives poor results - the
finite size effects are too strong to observe any power law in the
double logarithmic plot in fig.4. The second diagonal perpendicular
to the dipole moment however, though still with strong influence from
the boundaries, allows to fit some part of the 'bulk' data with
a power law (solid line in fig.4). The resulting exponent is 1.85,
and considering the small system size, quite close to the expected
value 2.

In conclusion we can say, that acoustic breathers can be obtained
for finite two-dimensional lattices up to numerical precision.
The symmetry is the one expected from general argumentations.
The dc components (and thus the strain) decay much slower than
the exponentially decaying ac components of the breather, and a
fit along one of the diagonals of the surprisingly small system
under study yields a power law with an estimated exponent of 1.85 to
be compared with the exponent of 2 which follows from the assumption
that the strain field has dipole symmetry.

These results should support the expectations that breathers can
exist in real crystals. Moreover at any finite temperature excited
breathers will decay after some time. Since they are accompanied
by a strain field, those strain fields will be dispersed in the
form of low-lying acoustic modes after the decay of a breather.
Thus breathers can act as an efficient energy transfer from high-lying
excitations into low-lying acoustic phonons.    

\newpage

\bibliography{db,kink}

\newpage

\noindent
FIGURE CAPTIONS
\\
\\
\\
Fig.1: 
dc displacements of a breather as a function of the lattice vector $l$.
Parameters as given in the text.
\\
\\
Fig.2:
Zoom of Fig.1 in the breather center.
\\
\\
Fig.3:
Absolute value of the strain of the breather solution of Fig.1
as a function of the lattice vector $l$.
\\
\\
Fig.4:
Variation of the absolute value of the strain (Fig.3) along the
diagonals of the lattice on a double logarithmic plot. 
Open circles - (1,1) direction; filled squares - (-1,1) direction.

\end{document}